\documentclass[conference]{IEEEtran}
\usepackage{amsmath,epsfig}
\usepackage{xcolor}
\usepackage{amssymb}
\usepackage{algorithm,algcompatible,amsmath}
\usepackage{multirow}
\usepackage{siunitx}
\usepackage{booktabs}
\usepackage{algorithm} 
\usepackage{color, colortbl}
\usepackage{enumitem}
\usepackage{algpseudocode} 
\usepackage{amsthm}
\usepackage{comment}
\usepackage{cuted}
\usepackage{Symbol}
\usepackage{multicol}
\usepackage{cite}
\usepackage{graphicx}
\usepackage{float}
\newtheorem{proposition}{Proposition}

\newtheorem{remark}{Remark}

\newtheorem{assumption}{Assumption}

\ifCLASSINFOpdf
\else
\fi
\begin{document}
%
\title{Online Filtering over Expanding Graphs}

\author{\IEEEauthorblockN{Bishwadeep Das}
\IEEEauthorblockA{Multimedia Computing Group\\
Delft Institute of Technology\\
Email: b.das@tudelft.nl}
\and
\IEEEauthorblockN{Elvin Isufi}
\IEEEauthorblockA{Multimedia Computing Group\\
Delft Institute of Technology\\
Email: e.isufi-1@tudelft.nl}
}


%


\maketitle

\begin{abstract}
Data processing tasks over graphs couple the data residing over the nodes with the topology through graph signal processing tools. Graph filters are one such prominent tool, having been used in applications such as denoising, interpolation, and classification. However, they are mainly used on fixed graphs although many networks grow in practice, with nodes continually attaching to the topology. Re-training the filter every time a new node attaches is computationally demanding; hence an online learning solution that adapts to the evolving graph is needed. We propose an online update of the filter, based on the principles of online machine learning. To update the filter, we perform online gradient descent, which has a provable regret bound with respect to the filter computed offline. We show the performance of our method for signal interpolation at the incoming nodes. Numerical results on synthetic and graph-based recommender systems show that the proposed approach compares well to the offline baseline filter while outperforming competitive approaches. These findings lay the foundation for efficient filtering over expanding graphs.
\end{abstract}

%
\IEEEpeerreviewmaketitle
\section{Introduction}
 Graph data-processing involves manipulating nodal information while obeying the underlying topology \cite{ortega_graph_2018}. Most works towards this end consider a topology with a fixed number of nodes but several applications involve expanding graphs with a growing number of nodes \cite{erdos_evolution_1961,barabasi_emergence_1999}. One such application is graph-based collaborative filtering with cold starters \cite{liu2020heterogeneous}, where we first incorporate a new user into the existing user-user graph and then process the data over this new topology to predict the ratings. Another application is predicting physical phenomena such as temperature or pressure at a new sensor joining an existing sensor network. 
 \par A central approach for solving these tasks is to use graph filters \cite{sandryhaila2013discrete,narang2013localized,coutino2019advances}. These are flexible, localized, parametric operators that combine neighboring node signals with a linear complexity in the number of nodes. Designing such filters requires knowledge of the full fixed topology but in settings where a stream of new nodes attach to the existing graph, i.e., the number of nodes keep growing, these methods require a re-training for each updated topology. To avoid this, we develop a filter update rule inspired by online machine learning that operates on the fly as new nodes attach to the graph \cite{shalev2012online,hazan2016introduction,orabona2019modern}.
 \par Pioneering works uniting online machine learning with graphs \cite{herbster2005online,herbster2006prediction,gu2013selective} have been focusing on online node classification on a static graph. For a dynamic setting, the two most relevant works are \cite{shen_online_2019} which uses random kernels to perform online learning on graphs using only the attachment pattern of the incoming node and \cite{chen2014semi}, which uses the inductive transference of graph filters. They classify a new node by extending the kernel feature similarities between new and existing nodes to the signal similarities at these nodes via a smoothing matrix. In \cite{shen_online_2019} the kernels are local, i.e., they rely on the location of the immediate neighbors and the predicted output does not rely on the graph data (e.g., signal values). Differently, graph filters rely on multi-resolution neighboring information and utilise the data over the topology to make predictions. This allows them to account for the influence of data over extended neighborhoods. In \cite{chen2014semi}, the existing nodes influence the incoming node through weights learned from kernel feature similarities to perform inference. Instead, we want to adapt the learning directly to the expanding graph. Differently, the work in \cite{dornaika2017efficient} estimates the attachment and uses it for for subsequent classification. The works in \cite{dasincoming2020,dasfiltering2020} deal with a stochastic attachment model for filter interpolation or a filter bank, but they are limited only to one incoming node an not a stream of nodes. Other works focus on mapping
 features to graph signals \cite{venkitaraman_recursive_2020}, node classification \cite{jian2018toward}, and dynamic embedding \cite{li2017attributed}. However, none of these works discuss graph filtering on continuously expanding graphs as a principled way to process graph data.

 \par We focus on online graph filter learning by accounting for the expanding graph topology. This yields filters that are localized over the growing graphs, inheriting the transference properties, and enjoy interpretability in terms of their frequency response. Our contributions are: $1)$ We propose a framework for the online learning of graph filters over continually expanding graphs; $2)$ We evaluate the learning for signal interpolation at each incoming node; $3)$ We corroborate the proposed model and provide interpretability via spectral analysis with the offline solution, while also comparing with the related literature for synthetic and real data. 
 \par The rest of this paper is as follows. Section 2 contains the problem formulation. Section 3 contains the online learning method. Section 4 contrasts our approach with others on synthetic and real data and Section 5 concludes the paper.
\section{Filtering Over Expanding Graphs}\label{Section PF}
Consider a starting graph $\ccalG_0=(\ccalV_0,\ccalE_0)$ of $N_0$ nodes, comprising a node set $\ccalV_0=\{v_1,\ldots,v_{N_0}\}$ and an edge set $\ccalE_0$. Let $\bbA_0\in\reals^{N_0\!\times\!N_0}$ be its adjacency matrix and $\bbx_0\in\reals^{N_0}$ the respective graph signal mapping each vertex in $\ccalV_0$ to $\reals$. Let $\{v_1,\ldots,v_{T}\}$ be a sequence of $T$ incoming nodes where $v_i$ arrives at time $t\!\!=\!\!i$, connects to the existing graph and stays attached for $t\geq i$. Under this rule, graph $\ccalG_t=\{\ccalV_t,\ccalE_t\}$ at time $t\!<\!T$ has $N_t\!\!=\!\!N_0\!+\!t$ nodes, adjacency matrix $\bbA_t\in\reals^{N_t\!\times\!N_t}$, and graph signal $\bbx_t\in\reals^{N_t}$. Node $v_{t+1}$ with signal $x_{t+1}$ attaches to $\ccalG_t$ with attachment vector $\bba_{t+1}\!=\![a_1,\ldots,a_{N_t}]^{\top}$, where each non-zero element of $\bba_{t+1}$ implies an edge between $v_{t+1}$ and $v\in\ccalV_t$. We consider each edge is directed from $v\in\ccalV_t$ to $v_{t+1}$. Such a setting is suited to inference tasks where existing nodes exert their influence over incoming ones, and not the other way around. The adjacency matrix $\bbA_{t+1}\!\in\!\reals^{N_{t+1}\!\times\!N_{t+1}}$ and graph signal $\bbx_{t+1}\!\in\!\reals^{N_{t+1}}$ have the respective\footnote{The undirected attachment case extends trivially to this update. As we focus only on inference at the incoming node, we assume the edges should not capture its influence over the existing vertices.} forms
    \begin{equation}
        \bbA_{t+1}=\begin{bmatrix}
        \bbA_t & \bb0 \\
        \bba_{t+1}^{\top} & 0 \\\end{bmatrix}~~~\textnormal{and}~~~
\bbx_{t+1}=\begin{bmatrix}
        \bbx_t \\
        x_{t+1} \\ \end{bmatrix}.
    \end{equation} 
    \par To process signal $\bbx_{t+1}$, we employ graph filters, which process graph data locally by combining successive shifts of an input signal over the topology. The output $\bby_{t+1}$ of a filter of order $K$ on $\bbx_{t+1}$ with shift operator $\bbA_{t+1}$ is 
    \begin{equation}\label{FIRop}
        \bby_{t+1}=\sum_{k=0}^{K}h_{t+1}^k\bbA_{t+1}^k\bbx_{t+1}
    \end{equation}where $h_{t+1}^k$ is the weight given to the $k$th shift $\bbA_{t+1}^k\bbx_{t+1}$. The vector $\bbh_{t+1}\!=\![h_{t+1}^{0},\ldots,h_{t+1}^{K+1}]^{\top}$ contains the filter parameters at time $t\!+\!1$. By substituting the $k$th adjacency matrix power
    \begin{equation}
        \bbA_{t+1}^k=\begin{bmatrix}
        \bbA_t^{k} & \bb0 \\
        \bba_{t+1}^{\top}\bbA_t^{k-1} & 0 \\\end{bmatrix}
    \end{equation} in \eqref{FIRop}, we can write the filter output
    \begin{equation}
        \bby_{t+1}=\begin{bmatrix}
        \sum_{k=0}^Kh_{t+1}^{k}\bbA_t^{l}\bbx_t \\
        \bba_{t+1}^{\top}\sum_{k=1}^Kh_{t+1}^{k}\bbA_t^{l-1}\bbx_t\\\end{bmatrix}.
    \end{equation}
    The output at $v_{t+1}$ is $[\bby_{t+1}]_{N_{t+1}}\!\!=\!\bba_{t+1}^{\top}\sum_{k=1}^Kh_{t+1}^{k}\bbA_t^{k-1}\bbx_t$.
    \par We learn the filter coefficients from a training set  $\ccalT=\{v_t,x_t,\bba_t\}_{t=1:T}$ where each element of $\ccalT$ comprises an incoming node $v_t$, its signal $x_t$, and its attachment vector $\bba_t$. We measure our performance at time $t\!+\!1$ relative to the signal $x_{t+1}$, revealed online, through a loss function $l_{t+1}([\bby_{t+1}]_{N_{t+1}},x_{t+1})$. Since the output is a function of the filter parameter, we consider the loss as $l_{t+1}(\bbh_{t+1},x_{t+1})$. The cumulative loss over the incoming node sequence is $L_{T}\!=\!\sum_{t=1}^{T}l_{t}(\bbh_{t},x_{t})$, where $\bbh_t$ is the filter generated by the online algorithm at time $t$.
    \par\smallskip\noindent\textbf{Problem statement.}\textit{ Given $(\ccalG_0,\bbA_0)$, signal $\bbx_0$, and the training set $\ccalT$, our goal is to predict a sequence of fixed-order FIR graph filters $\{\bbh_t\}_{t=1:T}$ generated online for minimizing a sequence of loss functions $l_t$ at the incoming nodes.}
\par\noindent We solve this with the tools of online machine learning, specifically online convex optimization to update the filter coefficients via online gradient descent (OGD) and characterize the regret of the approach relative to the optimal batch filter.
\section{Online Filter Learning}
Online filter learning implies updating parameters $\bbh_{t}$ based on $\bbh_{t-1}$ and the new node information (i.e., connections $\bba_{t}$ and signal $\bbx_{t}$). Focusing on signal reconstruction at the incoming node $v_t$, the loss function takes the form
\begin{equation}\label{instant loss}
l_t(\bbh_t,x_{t})=\frac{1}{2}(\bba_{t}^{\top}\bbA_{x,t-1}\bbh_t-x_{t})^2+\mu||\bbh_t||_2^2
\end{equation}
where $\bbA_{x,t-1}\!\!=\!\![\bb0,\bbx_{t-1},\bbA\bbx_{t-1},\ldots,\bbA_t^K\bbx_{t-1}]$ contains the shifted signals. The term $\mu||\bbh_t||_2^2$ acts as a regulariser to ensure there is no overfitting and $\mu>0$ is the respective weight. We perform the online update via gradient descent, i.e.,
\begin{equation}\label{online update}
    \bbh_{t+1}=\bbh_t-\eta_t\nabla_t
\end{equation}where the gradient $\nabla_{t}$ has the expression
\begin{equation}
 \nabla_t=(\bba_{t}^{\top}\bbA_{x,t-1}\bbh_t-x_{t})\bbA_{x,t-1}^{\top}\bba_t+2\mu\bbh_t   
\end{equation}
 and $\eta_t\!\!>\!\!0$ is the learning rate at time $t$. The loss is differentiable, convex in $\bbh_t$, and Lipschitz. The latter can be seen from the gradient expression, whose first term $\frac{1}{2}(\bba_{t}^{\top}\bbA_{x,t-1}\bbh_t-x_{t})\bbA_{x,t-1}^{\top}\bba_t$ is bounded as long as the graph signals have finite energy, the adjacency matrix is bounded (e.g., normalized adjacency matrix \cite{sandryhaila2014discrete}), and the filter $\bbh$ belongs to a closed set $\ccalH$ such that $||\bbh||_2\leq L_h$. Let the first term be bounded by a constant $C$. This makes the function Lipschitz with a factor of $(C+2\mu L_h)$, i.e,
$||\nabla_t||_2\leq(C+2\mu L_h)$ for all $t$. We denote the constant $(C+2\mu L_h)$ as $L$.
\par To measure the performance of the online algorithm w.r.t. the offline batch filter $\bbu$ computed upon observing the entire sequence, we pursue a standard regret analysis. We define the static regret of our algorithm w.r.t. the batch filter $\bbu$ over a $T$-length sequence as
    \begin{equation}\label{regret basic}
        R_{T}(\bbu)=\sum_{t=1}^{T}l_t(\bbh_t,x_{t})-l_t(\bbu,x_{t}).
    \end{equation}The regret measures how much worse off we are relative to an offline filter $\bbu$. We usually want the regret to grow at a sub-linear rate with time \cite{shalev2012online}. This implies $\frac{1}{T}\lim\limits_{T \to\infty}R_{T}(\bbu)=0$, i.e., the average performance of the algorithm asymptotically approaches the average performance of the offline filter $\bbu$. We can bound the regret of our algorithm $R_T(\bf u)$ as follows.
\begin{proposition}
Given the sequence of $L$ Lipschitz losses $l_t(\bbh_t,x_{t})$ [cf. \eqref{instant loss}] with $t\!=\!1\!,\ldots,\!T$ and a fixed gradient step size $\eta$ [cf. \eqref{online update}], the static regret $R_T(\bbu)$ [cf. \eqref{regret basic}] for the online algorithm generating filters $\{\bbh_t\}\in\ccalH$ relative to any other filter $\bbu\!\in\!\ccalH$ is upper bounded as
\begin{equation}\label{regret}
    R_T(\bbu)\leq \frac{||\bbu||_2^2}{2\eta}+\frac{\eta}{2}L^2T
\end{equation}
\end{proposition}
\textit{Proof}. The proof is similar to the regret for OGD [Thm. 2.13,\cite{orabona2019modern}].\qed\\
The bound holds with strict equality when $\bbu$ is the optimal batch filter. It is also a function of the step size $\eta$ and attains a minimum value of $||\!\bbu\!||_2L\sqrt{T}$ for $\eta=\frac{||\!\bbu\!||_2}{L\sqrt{T}}$. However, such an step size is unattainable in the online setting as it depends on the batch solution $\bbu$. As we focus more on filters and expanding graphs, we do not optimize the regret bound. However, there are approaches allied with OGD which do the same. Interested readers can refer to \cite{orabona2019modern}.
\par\smallskip\noindent\textbf{Computational complexity.}
Let us consider graph $\ccalG_0$ having $M_0$ edges and that each incoming node forms $P$ edges w.l.o.g. Graph $\ccalG_t$ has $M_0+Pt$ edges. The complexity of obtaining the filter output at time $t$ is $\ccalO((M_0+Pt)K)$, where $K$ is the filter order \cite{coutino2019advances}. The gradient and filter update step both share the same complexity. Thus at time $t$ the online algorithm has a complexity of order $\ccalO((M_0+Pt)K)$. The total complexity over a sequence of length $T$ summed over $t=1,\ldots,T$ is of the order  $\ccalO((M_0T+PT^2)K)$.
\begin{remark}
We discuss the static regret concerning a fixed offline filter $\bbu$, available once all the nodes in the sequence are observed. For our attachment method, this is the same as training the filter at all incoming nodes in the final graph $\ccalG_T$. One can alternatively consider the dynamic regret, where at each time instant we would have an optimal $\bbu_t$. Here, the online algorithm can be evaluated on how efficiently it tracks the optimal solution at each time step \cite{simonetto2016class}.\qed
\end{remark}
\section{Numerical Results}
We corroborate the proposed approach and compare it with both heuristic baselines and state-of-the-art alternatives across synthetic and real data. These comparisons are focused to answer the following research questions:
\par\noindent \textbf{RQ1.} \textit{How far off from the offline/ batch solution is the proposed method?}\\
We observe how much we lose in performance over the filter solved with complete knowledge of $\ccalT$. We compare with:
    \begin{enumerate}[label=\arabic*., start = 1]
    \item \textit{Batch Solution} (\textbf{BS}): This is the solution given by
    \begin{equation}
        \bbu=\underset{\bbh\in\reals^{K\!+\!1}}{\text{argmin}}\sum_{t=1}^T(\bba_{t}^{\top}\bbA_{x,t-1}\bbh-x_{t})^2+\mu_b|\!|\!\bbh|\!|_2^2
    \end{equation}
    \end{enumerate}which has a closed form least squares expression for $\mu_b\!\!>\!\!0$.
\par\noindent\textbf{RQ2.} \textit{How does online filtering for interpolation compare to online random kernel-based prediction?}\\
We compare with the state-of-the-art alternative:
\begin{enumerate}[label=\arabic*., start = 2]
\item \textit{Online Kernel Learning} (\textbf{OKL})\cite{shen_online_2019}: We consider a Gaussian kernel with variance $\sigma^2=\{1,10\}$. \end{enumerate}
\par\noindent\textbf{RQ3} \textit{How does online filtering for interpolation compare to pre-trained / inductive filtering?}\\
Following the inductive bias capabilities of graph filters \cite{gama2020graphs}, we want to contrast online learning to a filter pre-trained over $\ccalG_0$ and deployed on each $\ccalG_t$. Depending on the availability of signals at node $v_t$, we considered the alternatives
    \begin{enumerate}[label=\arabic*., start = 3]
        \item \textit{Inductive Transfer without features} (\textbf{IT1}): We learn a filter on $\ccalG_0$ using $\bbA_0,\!\bbx_0$ and deploy it on $\ccalG_1,\ldots,\ccalG_T$ to predict the signals at $v_1,\ldots,v_T$, respectively without retraining. We train the filter through a regularized least square problem \cite{sandryhaila2013discrete}. 
        \item \textit{Inductive Transfer with features} (\textbf{IT2})\cite{chen2014semi}: First, we learn the filter in \textbf{IT1}. The output at $v_t$ is predicted as $y_{t}=\bbm^{\top}\bbA_{x,t-1}\bbh$, where $\bbh$ is the pre-trained filter, $\bbm\in\reals^{N_t}$ is a smoothing vector which contains the relational dependencies between the signals at the existing and incoming nodes. This vector is obtained as $\bbk'(\bbK+\lambda\bbI)^{-1}$, where $\bbk$ contains the kernel similarities between the features of $v_t$ and $\ccalV_{t-1}$, and $\bbK$ the same between the nodes in $\ccalV_{t-1}$ \cite{chen2014semi}. We report the better of the performance between a linear and a Gaussian kernel with $\sigma^2={1,5,10}$.
    \end{enumerate}
We also compare with the standard baseline:
\begin{enumerate}[label=\arabic*., start = 5]
\item \textit{Weighted K-Nearest Neighbor Prediction} (\textbf{WKNN}): The predicted signal at $v_t$ is the weighted mean of the signal at the nodes to which $v_t$ attaches, with the corresponding edge weights normalized to have unit sum.
\end{enumerate}
We evaluate all approaches via the root Normalized Mean Square Error (rNMSE) over the sequence
\begin{equation}
    \text{rNMSE}=\sqrt{\frac{\sum_{t=1}^T(\hhatx_t-x_t)_2^2}{\sum_{t=1}^Tx_t^2}}
\end{equation}where $\hhatx_t$ and $x_t$ are the predicted and true signal at $v_t$, respectively. All filters have order $K\!\!=\!\!5$. Both the online methods (proposed and \textbf{OKL}) have parameters of equal dimension and have the same hyper-parameters $\eta$ and $\mu$ selected via grid search over $[10^{-6},10]$. For $\textbf{BS}$, $\textbf{IT1}$, and $\textbf{IT2}$, the regularisation hyper-parameter is chosen over $[10^{-4},10]$.
\definecolor{Gray}{gray}{0.9}
\begin{table*}[t]
\centering
\caption{\label{rNMSE synthetic}rNMSE and standard deviation of all approaches averaged over ten synthetic data-sets for each signal generation scenario.}
\begin{tabular}{c ||c c c c c||c c c c c||c c c c c}
\hline\hline
& \multicolumn{5}{c}{\footnotesize{Filter}} & \multicolumn{5}{c}{\footnotesize{WMean}} & \multicolumn{5}{c}{\footnotesize{Kernel}}\\
\hline
\footnotesize{Method} & \footnotesize{\textbf{Prop.}} & $\textbf{\footnotesize{BS}}$ & $\textbf{\footnotesize{OKL}}$ & $\textbf{\footnotesize{IT1}}$ & $\textbf{\footnotesize{WKNN}}$ & \footnotesize{\textbf{Prop.}} & $\textbf{\footnotesize{BS}}$ & $\textbf{\footnotesize{OKL}}$ & $\textbf{\footnotesize{IT1}}$ & $\textbf{\footnotesize{WKNN}}$ &  \footnotesize{\textbf{Prop.}} & $\textbf{\footnotesize{BS}}$ & $\textbf{\footnotesize{OKL}}$ & $\textbf{\footnotesize{IT1}}$ & $\textbf{\footnotesize{WKNN}}$\\\hline
\rowcolor{Gray}
\footnotesize{rNMSE} & 0.09 & $3\!\times\!10^{-4}$ & 0.82  & 0.26 & 0.27 &0.09 & 0.034 & 0.71 & 0.27 & 0 & 0.50 & 0.39 & 0.39 & 4.79 & 4.03\\
\footnotesize{sdev} & 0.03 & $1\!\times\!10^{-4}$ & 0.19 & 0.06 & 0.13 & 0.02& 0.007 & 0.2 & 0.07 & 0 & 0.05 & 0.04 & 0.04 & 0.38 & 0.47\\
\hline\hline
\end{tabular}
\label{synthetic de-noising}
\end{table*}
\begin{figure*}[t]
\centering
{\includegraphics[trim=30 210 50 230,clip,width=0.32\textwidth]{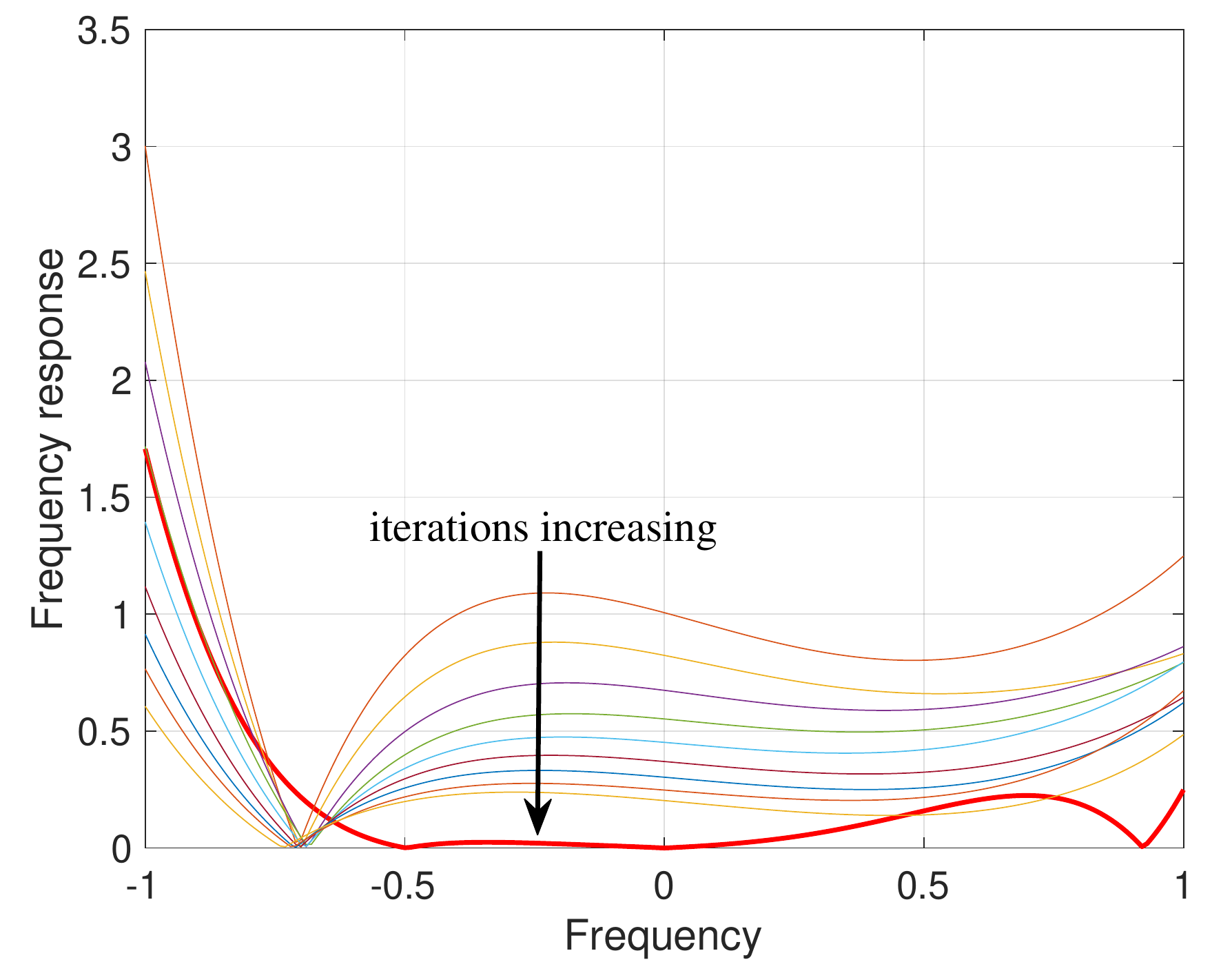}}%
\includegraphics[trim=30 210 50 230,clip,width=0.32\textwidth]{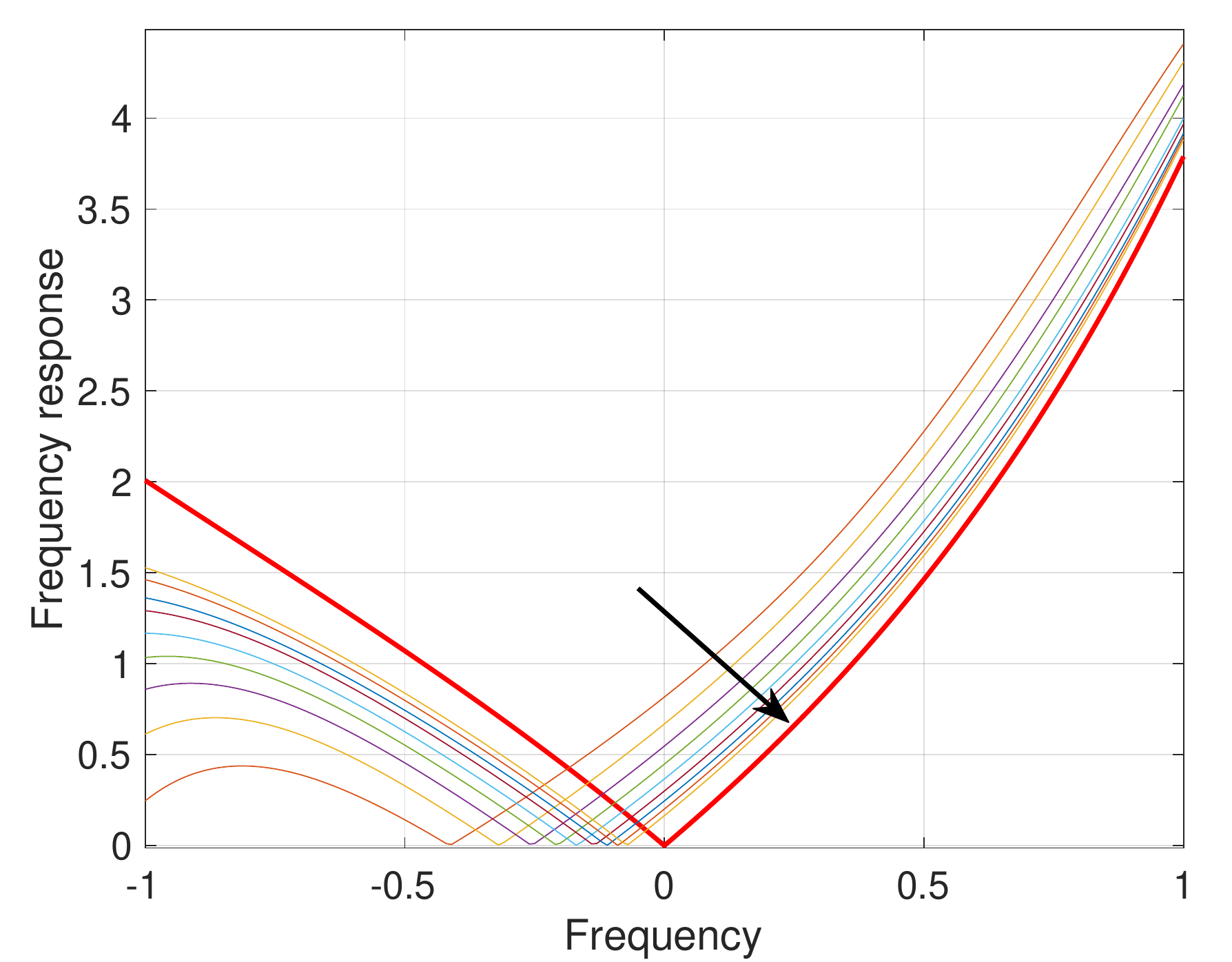}%
{\includegraphics[trim=30 210 50 230,clip,width=0.32\textwidth]{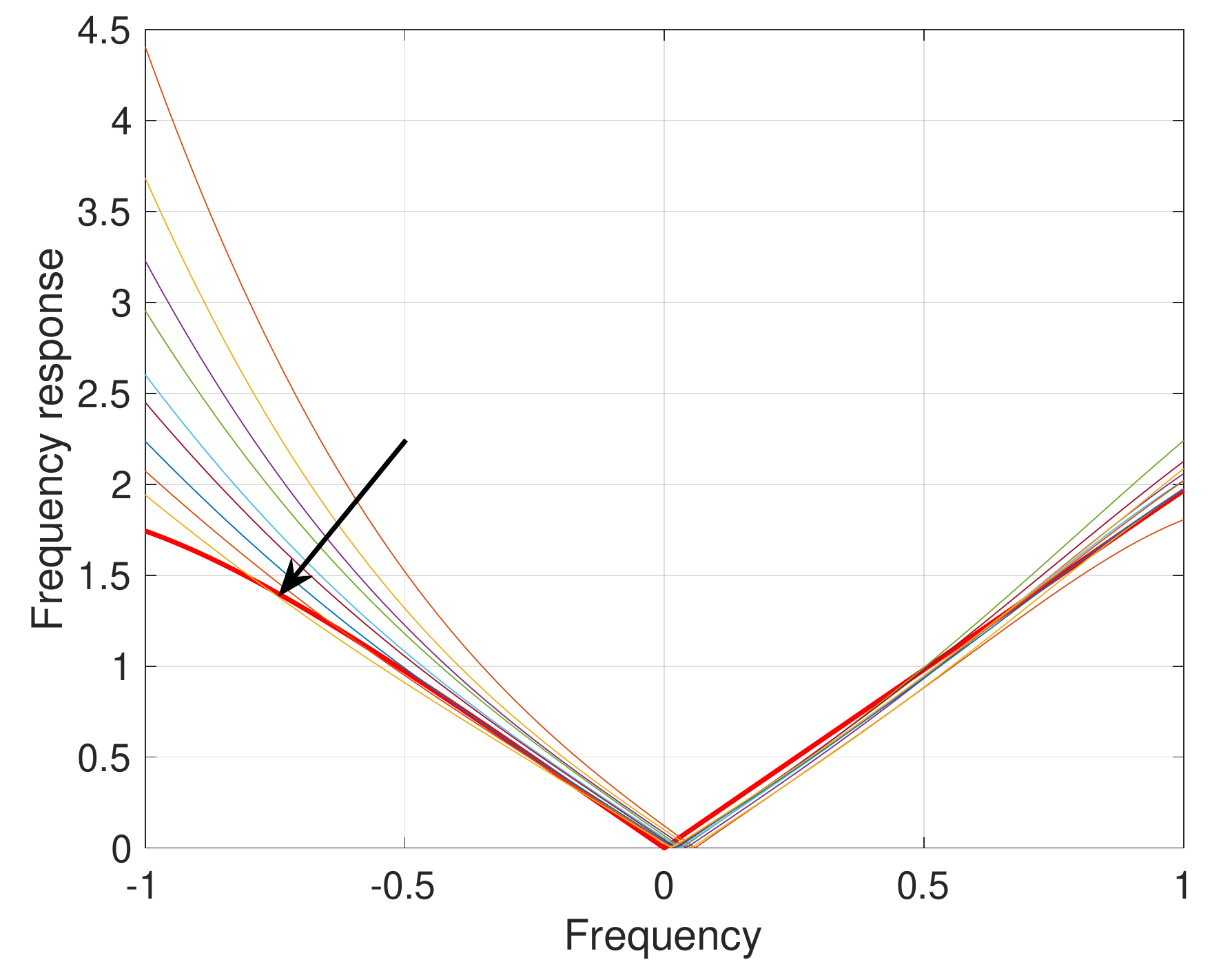}}%
\caption{Frequency response of the online filter (colored lines) every $100$ iterations compared with that of batch solution $\textbf{BS}$ (in bold red) for (left) \textit{Kernel}, (centre) \textit{Filter}, and (right) \textit{WMean} generated data.} 
\label{freq_plots} 
\end{figure*}
\subsection{Synthetic}
\noindent\textbf{Experimental setup.} We consider an Erdos-Renyi graph $\ccalG_0(N,p)$ with $N=100$ and $p=0.2$ \cite{erdos_evolution_1961}. We normalize the adjacency matrix by the largest absolute eigenvalue to stabilise the filter output as in \cite{sandryhaila2013discrete}. We construct the training set $\ccalT$ with each $v_t$ forming $P=5$ edges \textit{uniformly-at-random} relative to the existing graph $\ccalG_t$. Each edge has a weight equal to the median of the edge weights in $\ccalE_0$. To generate $x_t$ at $v_t$, we consider three options:
\begin{enumerate}
\item \textit{Kernel}, where all $x_t$s are generated following \cite{shen_online_2019}, which favors kernel solutions.
\item \textit{Filter}, where $x_t$ is generated using a pre-trained filter with $K=5$ on $\ccalG_0$ using data at $80$ percent of its nodes, which favors filtering solutions. \item \textit{WMean}, where $x_t$ is the $\textbf{WKNN}$ prediction, which is a neutral choice.
\end{enumerate}
Our motivation here is to see how each of the methods compares in settings that may or may not be suitable for it. We iterate all methods over $20$ initial graphs $\ccalG_0(N,p)$, with $T=1000$ for each realization. Since $v_t$ has no features, we consider $\textbf{IT1}$ for $\textbf{RQ3}$, whose performance is averaged over $50$ sets of sampled signals for pre-training.
\par\smallskip\noindent\textbf{Results.} Table \ref{rNMSE synthetic} showcases the rNMSE along with the standard deviation for all approaches averaged across all graphs in all three scenarios. Concerning $\textbf{RQ1}$, the online solution underperforms \textbf{BS} as expected with the gap being the most for \textit{Filter} data. This is because the setting is biased towards filtering. For $\textbf{RQ2}$, the proposed outperforms $\textbf{OKL}$ except in the case for \textit{Kernel} data, which favours $\textbf{OKL}$ but still the filtering solution is close to it. This indicates the advantage of graph filters that incorporate higher-order node interactions through successive shifts, as opposed to the kernel model which relies only on the incoming attachment. For $\textbf{RQ3}$, the proposed method outperforms $\textbf{IT1}$ as online approaches adapt to the incoming data and topology change better than pre-trained filters which are biased to one signal and topology. Note that the zero error obtained by $\textbf{WKNN}$ for \textit{WMean} data is due to them being the same method, and is of no importance.
\definecolor{Gray}{gray}{0.9}
\vspace{-6mm}
\begin{table}[h]
\centering
\caption{Normalized Regret for all three data generations}
\begin{tabular}{c|c c c c }
\hline\hline
\footnotesize{Method} & $\text{\footnotesize{Kernel}}$ & $\text{\footnotesize{Filter}}$ & $\text{\footnotesize{wMean}}$ & $\footnotesize{\text{MovieLens}}$ \\
\rowcolor{Gray}
\footnotesize{$R_T(\bbu)/T$} & 0.29  & 0.0021  & 0.0043 & 0.12 \\
\hline\hline
\end{tabular}
\label{normalized regret synthetic}
\vspace{-3mm}
\end{table}
\par Fig. \ref{freq_plots} showcases how the online filter processes data over the existing graph in the frequency domain. We plot the frequency response of the online filter every $100$ iterations and that of the batch over the frequencies $[-1,1]$ following the frequency response interpretation w.r.t. the adjacency matrix \cite{sandryhaila2014discrete}. Frequencies approaching $\!-\!1\!$ and $\!1\!$ are the high and low frequencies, respectively. The batch frequency response is in bold red. For \textit{Filter} and \textit{wMean} data, the online frequency response tends towards the batch response. Table \ref{normalized regret synthetic} contains the normalized regret and we see that for \textit{Filter} and \textit{wMean}, the values are quite low for $T=1000$, as suggested by the respective frequency plots. For \textit{Kernel} data, the online filter approaches the batch response for the middle and low frequencies while attenuating the high frequencies more. This could be because of an insufficient number of incoming nodes. This gap is also reflected in the regret value in Table \ref{normalized regret synthetic}. 
\subsection{Collaborative filtering}
In this section, we predict ratings for a sequence of incoming users using collaborative filtering. We considered the Movielens100K data-set, comprising $100,000$ ratings provided by $943$ users over $1152$ items \cite{harper2015movielens} with at least ten ratings. We use graph collaborative filters for this task \cite{huang_rating_2018}.
\par\noindent\textbf{Experimental setup}. We consider $N_0\!=\!500$ starting users at random and treat the remaining $443$ as online. We build the $15$-NN directed user-user graph $\ccalG_0$ with normalized adjacency matrix $\bbA_0$ for the starting users using the cosine similarity of their rating vectors\footnote{All online methods performed better at this range of nearest neighbors.}. For each new user $v_t$, we use $50$ percent of the ratings to build $\bba_t$, which has $15$ edges connected to nearest neighbors via cosine similarity. We predict the remaining ratings. We evaluated all methods over $20$ realizations of the basic setup. Since each incoming user has available features, i,e, a subset of ratings, we compare also with $\textbf{IT2}$. We train this filter over the same number of existing ratings as the number of online ratings.
\definecolor{Gray}{gray}{0.9}
\begin{table}[h]
\centering
\caption{rNMSE of all approaches averaged over $20$ realizations for user rating prediction in Movielens100K}
\begin{tabular}{c|c c c c c c}
\hline\hline
\footnotesize{Method} & \footnotesize{\textbf{Prop.}} & $\textbf{\footnotesize{BS}}$ & $\textbf{\footnotesize{OKL}}$ &
$\textbf{\footnotesize{IT2}}$ & $\textbf{\footnotesize{WKNN}}$ \\
\rowcolor{Gray}
\footnotesize{rNMSE} & 0.56 & 0.55 & 0.69 & 0.6 & 0.61\\
\footnotesize{sdev} & 0.02 & 0.006 & 0.006 & 0.03 & 0.008\\
\hline\hline
\end{tabular}
\label{table rating prediction}
\end{table}
\par\noindent\textbf{Results}.
Table \ref{table rating prediction} showcases the rNMSE of all approaches for this experiment. Concerning $\textbf{RQ1}$, the proposed is close to $\textbf{BS}$, while regarding $\textbf{RQ2}$ and $\textbf{RQ3}$, the proposed does better than $\textbf{OKL}$ and $\textbf{IT2}$, by improving the rNMSE by up to $18.8\%$ and $8.2\%$, respectively.
\par
\begin{figure}[t]
\centering
\includegraphics[trim=0 190 0 200,clip,width=0.3\textwidth]{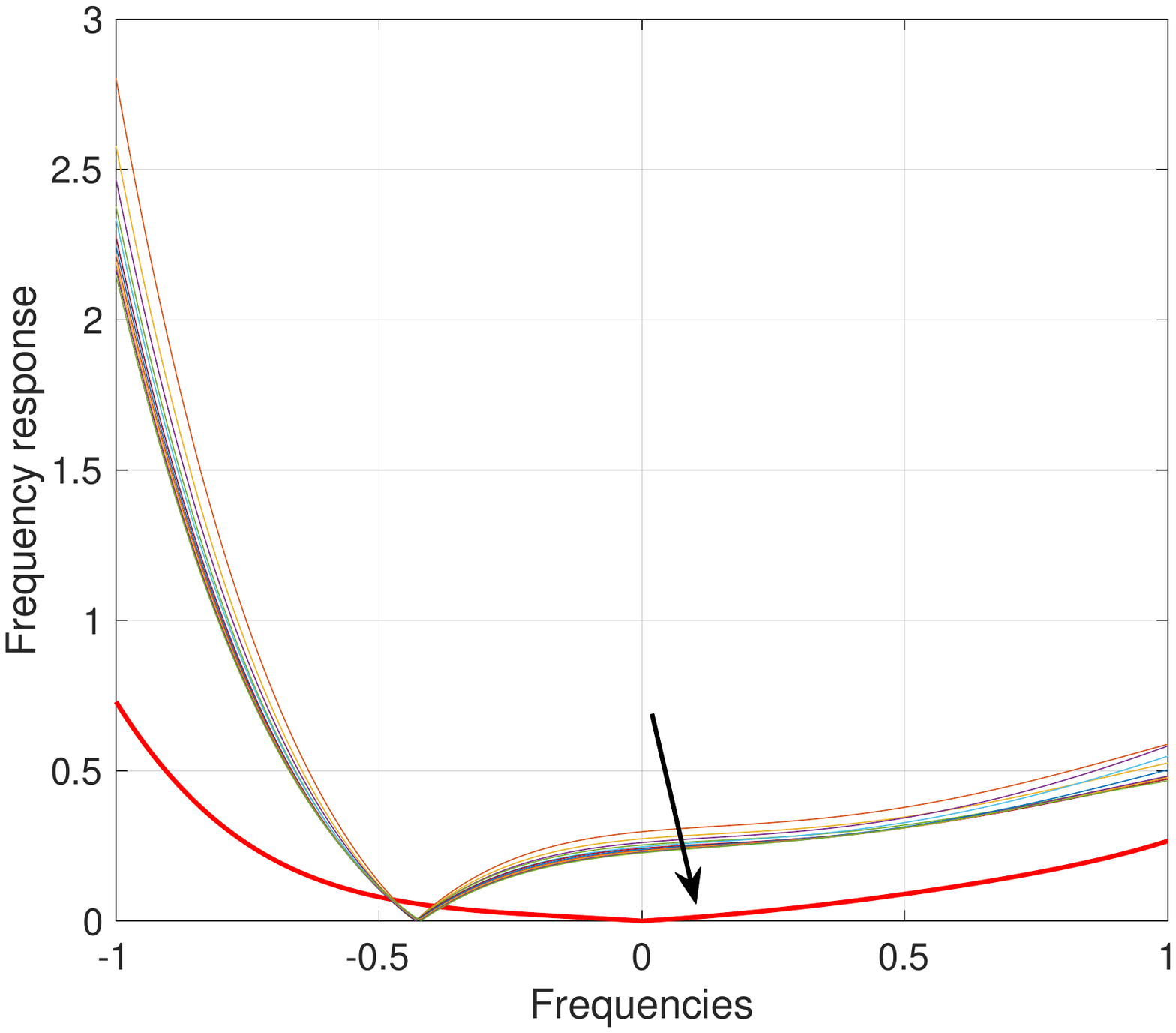}%
\caption{Frequency response of the online filter every $2000$ updates compared with that of $\textbf{BS}$ (bold red) for Movielens-100K} 
\label{freq_plots_recsys} 
\end{figure}

 Fig. \ref{freq_plots_recsys} plots the frequency response of the online filter every $2000$ steps over roughly $23,000$ samples in one run of an experiment. It also contains the response of $\textbf{BS}$. Despite achieving an rNMSE close to $\textbf{BS}$, the frequency responses are not close, with a normalized regret of $0.12$ [cf. Table \ref{normalized regret synthetic}]. The online filter has a higher amplitude for higher frequencies than the batch filter. For recommender systems, this implies a greater diversity in the recommendation (i.e., prioritizing good ratings to diverse items and not only to a few popular ones)\cite{isufi2021accuracy}. To test this, we evaluate the average  ten item aggregated diversity $\text{AD@10}$ for both approaches. This shows what fraction of the total items appear in a list containing the top ten recommendation over all users, with a higher value showing a better diversity. We evaluate this online for the proposed and with the batch filter for all incoming users. The proposed and \textbf{BS} have average $\text{AD@10}$ as $0.17\pm0.01$ and $0.12\pm0.01$, respectively. The proposed achieves a higher value, thus showing that the online filter promotes whioe retaining a similar rNMSE.
\section{Conclusion}
We proposed online graph filtering for expanding graphs over a sequence of incoming nodes. With each incoming node, the filter processes the existing signal to make a prediction. We update the filter parameters based on an online gradient descent update over the loss. Numerical results over synthetic and real data for signal interpolation at the incoming node show that the proposed approach is competitive w.r.t. the offline filter in the rNMSE while also exhibiting similar behaviour in the frequency domain, also providing more diverse recommendations. Results also show the improvement in rNMSE over online random kernels and inductive filtering. For future work, we will consider the scenario where the existing graph data obeys a time varying model and adapt the online algorithm to be graph-time aware.



%



\bibliography{ref}
\bibliographystyle{unsrt}

\end{document}